\documentclass[aps,prc,showpacs,amsmath,amssymb]{revtex4}
\usepackage{epsfig}

\newcommand{\acal}{{\cal A}}

\begin{document}

\title{Influence of the nucleon spectral function\\
in photon and electron induced reactions on nuclei}

\author{J. Lehr and U. Mosel}

\affiliation{Institut f\"ur Theoretische Physik, Universit\"at Giessen\\ 
            D-35392 Giessen, Germany}

\date{\today}

% ----------------------------------------------------------------
\begin{abstract}
We study the influence of the nucleon spectral function on $\eta$ photo- and
electroproduction on nuclei. Besides kinematical effects due to
groundstate correlations, also a modification of the $S_{11}(1535)$
decay width is taken into account, which is caused by the possible
decay into nucleons with mass smaller than the pole mass in the medium.
Hence, resonances with masses below the free $N\eta$ threshold can 
contribute to $\eta$ production.
\end{abstract}
% ----------------------------------------------------------------
\maketitle

\section{Introduction}

Nucleons inside nuclei are commonly described in terms of quasi particles
with an effective mass connected to the mean-field potential. However, it 
also known that such strongly interacting particles obtain a finite width 
\cite{kad-baym} from processes that correspond in lowest order (beyond the 
mean-field approximation) to collision reactions. The so-called collisional 
broadening is therefore a feature that should be accounted for in nuclear 
reactions where the nucleons frequently collide with each other or with
other particles. 
The nucleons in the nuclear groundstate are subject 
to a broadening due to short-range correlations (see e.g.
\cite{benhar,cda,corr1,corr2}). In inclusive reactions such as electron 
scattering off nuclei these effects have been taken into account and proven 
to have influence (e.g. \cite{benhar_scatt}). In \cite{sibi1,sibi2} 
photon induced $K^+\Lambda$ production and proton induced production of
heavy mesons $K^+,\rho,\omega,\phi$ on carbon has been studied within the
framework of folding models showing also an influence in more 
exklusive reactions.
In the energy regime where nucleon resonances play an important role
there is another modification that should be taken into account: In the
vacuum, it is known that the decay width of resonances into channels with
unstable mesons (e.g. $R\to N\rho$) is strongly influenced by the 
meson spectral function. If the rho meson was a stable particle with a
mass of 770 MeV, a $D_{13}(1520)$ resonance with a mass of 1.524 GeV could
not decay into $N\rho$. However, the branching ratio into this channel is
in the range of 15-25\% \cite{pdg}. Therefore, also the decay width of the
nucleon resonances into channels with nucleons should change when the
nucleon spectral function deviates from a delta function in the medium.

In the framework of transport models, theoretical concepts 
for the offshell propagation of nucleons have been developed 
\cite{effe_offshell,cassjuch,leupold}, which was a prerequisite for the
treatment of offshell nucleons in heavy-ion reactions.
We are, therefore, now in a position to be able to combine the aspects of the 
groundstate correlations and the nucleon spectral function in the FSI as well 
as the propagation to study more exclusive reactions with electromagnetic 
probes. In this work, we consider $\eta$ photo- and electroproduction on 
nuclei 
$(T=0)$ in the second resonance region and study the influence of the nucleon 
spectral function. For these energies, this channel is particularly 
interesting, because of the strong connection between the $\eta$ meson and the 
$S_{11}(1535)$ resonance.
In particular we will focus on the sensitivity of such reactions on the 
groundstate correlations and the modification of the $S_{11}(1535)$ decay
width.
For our calculations on nuclei we use a semi-classical BUU transport model,
which we already used for the calculation of these processes using the
onshell approximation for the nucleons \cite{photoeta,electroeta}.

We start in Sec. \ref{sec:nucl-spec-fct} with a discussion of the nucleon
spectral function used in our calculations. In Sec. \ref{sec:s11-dec}
we describe how the decay widths are modified by the nucleon spectral
function in the medium, followed by some details about the the
calculation of the cross sections on nuclei and the BUU model in Sec.
\ref{sec:el-react} and \ref{sec:buu-model}.
In Sec. \ref{sec:results} we show our results on $\eta$ photo- and 
electroproduction on calcium.

%%%%%%%%%%%%%%%%%%%%%%%%%%%%%%%%%

\section{The nucleon spectral function}
 \label{sec:nucl-spec-fct}

\begin{figure}
\begin{center}
\includegraphics[width=15cm]{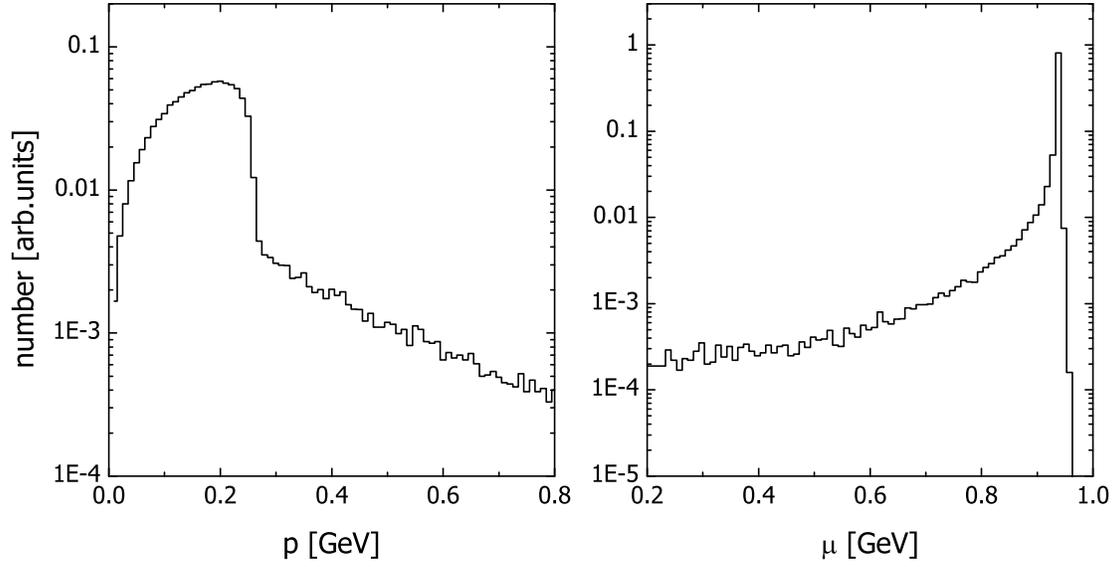}
\end{center}
\vspace{-0.5cm}
\caption{Momentum and mass spectra for the nucleons inside calcium 
including groundstate correlations using local density approximation.}
 \label{fig:mom-mass-spec}
\end{figure}

The influence of the nucleon spectral function is twofold: On the one
hand, there are correlations between the nucleons in the nuclear
groundstate leading to a finite width. The nucleon spectral function
in nuclear matter was calculated in \cite{corr1,corr2} using transport 
theoretical methods. The width was obtained by calculating collision widths.
There it was found that for a description of the 
spectral function properties the knowledge of the overall strength of
the nucleon interaction is already sufficient. Determining this parameter
from more sophisticated many-body models \cite{benhar}, both momentum
cuts of the spectral function and the momentum distribution are in very
good agreement with such approaches.
We have performed the same calculations for different densities. Using
local density approximation, we were then able to apply our nucleon
spectral function also to the groundstate of finite nuclei. A similar
procedure was used in \cite{stringari} to obtain the momentum distribution
in finite nuclei.
In Fig. \ref{fig:mom-mass-spec} we show the mass and momentum spectra in a
calcium nucleus, determined by the spectral function. It is seen than the mass 
spectrum shows a strong asymmetry with respect to the still dominating onshell 
peak; most nucleons off the
mass shell have masses $\mu<m_N$. This can be easily understood if we
consider for simplicity nuclear matter. In the groundstate all states
with energy smaller than the Fermi energy $E_F$ are occupied. The nucleons
with pole mass $m_N$ obey the condition $E=\sqrt{m_N^2+p^2}$. 
The region in the $(E,p)$ plane, where $\mu>m_N$, i.e. 
$E>\sqrt{m_N^2+p^2}$, is confined by the onshell curve, the $E$ axis and 
the condition $E\le E_F$ and is therefore only a small fraction of the
whole area $E\le E_F$.
\medskip

\begin{figure}
\begin{center}
\includegraphics[width=11cm]{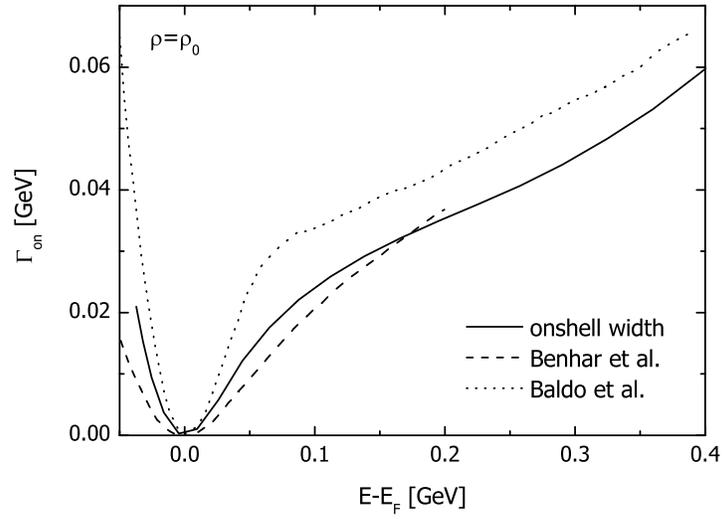}
\end{center}
\vspace{-0.5cm}
\caption{Onshell width of the nucleons calculated at nuclear matter
 density $\rho_0$ as a function of the energy. The dashed and dotted 
curves show the result from Benhar et al. \cite{benhar} and Baldo et
al. \cite{baldo}.}
 \label{fig:onshell-width}
\end{figure}

In addition, the nucleons undergo collisions in the course of the FSI. 
This also gives rise to a finite width for nucleons with energies above the
Fermi energy. These widths are determined by evaluating 
the collision rates for the reactions involving nucleon as incoming particles 
for energies above the Fermi energy (at $T=0$ the widths vanish otherwise
due to Pauli blocking).
For elastic nucleon-nucleon scattering $N+N_2\to N_3+N_4$ e.g.
we obtain \cite{effe_abs}
\begin{equation}
 \label{eq:collrate}
  \Gamma = g\int {d^3p_2\over(2\pi)^3}\,d\mu_2\,d\mu_3\,d\mu_4\,d\Omega\, 
    v_{\textrm{rel}}{d\sigma_{\textrm{c.m.}}\over d\Omega} \acal_2f_2
   \acal_3(1-f_3)\acal_4(1-f_4)
\end{equation}
where $v_{\textrm{rel}}$ is the relative velocity of the incoming particles
and $g$ is the spin-isospin degeneracy factor. The spectral function
$\acal$ is given by \cite{effe_dilep}
\begin{equation}
  \label{eq:spec-fct}
  \acal_N(\mu_N,p_N,\rho)={2\over\pi}{\mu_N^2\Gamma(p_N,\rho)\over
    (\mu_N^2-m_N^2)^2+\mu_N^2\Gamma^2(p_N,\rho)}.
\end{equation}
The widths for the FSI reactions are therefore calculated on the same footing
as for the groundstate correlations. 
For other collisional processes, equivalent expressions can be written down. 
Note that the cross section for the nucleon reactions enters Eq.
(\ref{eq:collrate}). Since we make use of vacuum cross sections 
\cite{effe_dilep}, we only calculate the nucleon spectral function for the 
FSI on the nucleon mass shell. Therefore, the resulting width only depends on 
density and momentum, but not independently on the nucleon energy.
For the nucleons, the total width is obtained by calculating the collision 
rates for the processes $NN\to NN$, $NN\to NR$, $NN\to \Delta\Delta$ and
$NN\to NN\pi$, which are explicitly implemented in our transport model
(for details see Ref. \cite{effe_abs}). Other 
contributions can be neglected, because the
multiplicities of the nucleon scattering partners are negligible. For
larger energies, when more inelastic multi-particle channels open, we use in
Eq. (\ref{eq:collrate}) the parametrization for the total cross section 
from \cite{pdg} and evaluate the collision width by omitting the 
Pauli blocking factors for the outgoing fermions.
In Fig. \ref{fig:onshell-width} we compare 
the width in the nuclear matter rest frame for density $\rho_0$ as a function 
of energy with the onshell  widths obtained from the many-body results of 
\cite{benhar,baldo}. In the considered energy range, the agreement is very 
good. 

\begin{figure}
\begin{center}
\includegraphics[width=11cm]{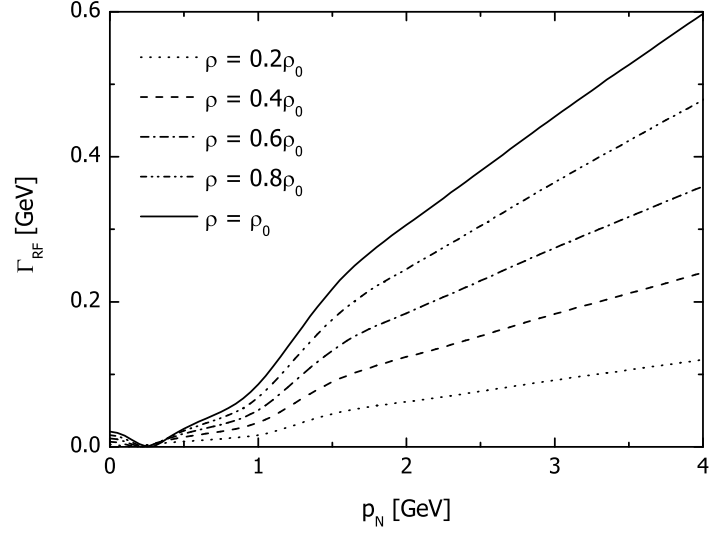}
\end{center}
\vspace{-0.5cm}
\caption{Onshell nucleon width as a function of the nucleon momentum in the
  lab frame for different densities, calculated in the rest frame of the
nucleon.}
 \label{fig:nwidth-dens}
\end{figure}

We have calculated the nucleon width as a function of momentum.
In Fig. \ref{fig:nwidth-dens} the width in the
rest frame of the nucleon is shown for different density values as a function
of momentum.
The width exhibits the zero at the Fermi momentum and then increases with
growing inelasticity.
The connection of the width in the nucleon rest frame (RF) to that in the 
nuclear matter rest frame (LAB) is given by
\begin{equation*}
  \Gamma_{\textrm{RF}}={E\over\mu}\cdot\Gamma_{\textrm{LAB}}.
\end{equation*}
Therefore, $\Gamma_{\textrm{RF}}$ is larger than $\Gamma_{\textrm{LAB}}$.

It should be mentioned that the nucleon width arising from the FSI 
in Eq. (\ref{eq:collrate}) for nucleon energies above the Fermi energy $E_F$ is
calculated on the same footing as the width arising from the groundstate
correlations in \cite{corr1,corr2} for energies below $E_F$.
The difference is, that in \cite{corr1,corr2} we used an averaged in-medium
matrix element, whereas in Eq. (1) the matrix element is substituted by
the NN-vacuum cross section. Both widths, however, approach each other
because they vanish at $E= E_F$.

%%%%%%%%%%%%%%%%%%%%%%%%%%%

\section{$S_{11}(1535)$ decay width in medium}
 \label{sec:s11-dec}

In this Section we discuss how the nucleon spectral function influences
the in-medium decay of nucleon resonances. 
The strength of the effect is determined by the nucleon width. As 
discussed in Sec. \ref{sec:nucl-spec-fct}, this depends on the momentum of the
nucleons stemming from the resonance decay, and the density. There it was 
shown (see Fig. 3) that with increasing momentum, the in-medium width of the 
nucleon reaches values comparable to the rho width in vacuum (250 MeV).
Therefore, an appreciable effect of nuclear collisional broadening should 
also be seen. The question remains in which reactions resonances with such 
large momenta can be prepared.
Recently we have discussed $\eta$ photo- and electroproduction on nuclei
\cite{photoeta,electroeta} in the onshell limit for the nucleons.
In this case, the $S_{11}(1535)$ is the dominating degree of freedom for
eta production. Most detected etas stem from the decay of a $S_{11}$ 
produced in the elementary reaction $\gamma N$ or $\gamma^* N$. The 
resonance momentum depends on the momentum transfer in this reaction.
For virtual photons, the virtuality $Q^2$ can be chosen independently 
from  the photon energy, which allows for a preparation of resonances
with a certain fixed mass and different momenta: In photoproduction, a 
$S_{11}$ resonance with pole mass has a momentum of 0.78 GeV, in
electroproduction at $Q^2=3.6$ GeV$^2$ a momentum of 3.3 GeV.
Also resonances experience a collisional broadening \cite{effe_abs}.
For resonance momenta up to 1 GeV as encountered in photoproduction,
the collision width is rather small ($\sim$ 30 MeV at $\rho_0$) 
\cite{photoeta}. The calculation for larger resonance momenta of several
GeV by using collision rates is difficult because the cross section
$N S_{11}\to X$ in this kinematical regime is unknown. Also estimates based on
the total $NN$ cross section would not help because a broadening and/or
suppression of the electron-nucleus cross section caused by the collision
width depends on a reliable separation of $\Gamma_{\textrm{coll}}$ into
contributions arising from absorptive reactions and processes with a $S_{11}$ 
or other particles leading to $\eta$ production in the final state
\cite{d13paper}.
We therefore focus only on the modification of the in-medium decay width.
Since we are interested in $\eta$ production, we confine our considerations 
to a modification of the $S_{11}(1535)$ only. Moreover, only the most important
decay channels, $N\pi$ and $N\eta$, will be modified.
The vacuum resonance widths used in our BUU transport model are parametrized
in the same way as in \cite{manley,effe_dilep}, which for the resonance decay 
$R\to Nm$ is given by:
\begin{equation}
  \label{eq:manley-par}
  \Gamma_{R\to Nm}(\mu_R)=\Gamma_0{\rho_{Nm}(\mu_R)\over \rho_{Nm}(M_R)},
\end{equation}
where $\Gamma_0$ denotes the total decay width at the pole mass $M_R$. 
In the vacuum, the $\rho$-function for a decay into a stable nucleon and a 
stable meson reads:
\begin{equation}
  \label{eq:rho}
  \rho_{R\to Nm}(\mu)={p_{Nm}(\mu_R)\over\mu_R}B_l^2(p_{Nm}).
\end{equation}
Here $p_{Nm}$ is the c.m. momentum of the decay products, $\mu_R$ is the
resonance mass and $B_l$ is the Blatt-Weisskopf function for the relative
angular momentum $l$ of the decay products, which for low c.m. momenta behaves 
like $\propto p_{Nm}^l$, thereby ensuring the correct momentum dependence
of the width close to threshold \cite{manley}. In the case
of the $S_{11}(1535)$ we have $B_l=1$. The expression for $\rho_{Nm}$ is
close to the general expression for the width 
$d\Gamma\sim\vert{\cal M}\vert^2/\mu_Rd\Phi_2$ \cite{pdg}. The factor 
$p_{Nm}/\mu_R$ can be identified with the two-body phase space. The number 
$\rho_{Nm}(M_R)$ in the denominator in Eq. (\ref{eq:manley-par})
therefore is part of the matrix element which describes the coupling $R\to Nm$ 
and will be left untouched when we go to the medium.
For the decay of a resonance into a stable meson and a broad nucleon in 
the medium, the function $\rho_{Nm}$ in the numerator of Eq. 
(\ref{eq:manley-par}) is extended by performing an integration over the 
nucleon spectral function. This is actually the same as what is done in
\cite{manley} for the parametrization of the resonance decay in vacuum into 
a stable nucleon and an unstable meson, e.g. $R\to N\rho$.
Hence, we end up with the expression
\begin{equation}
  \label{eq:inmed-width}
  \Gamma_{S_{11}\to Nm}^*(\mu_R,p_R,\rho)={\Gamma_0\over\rho_{Nm}^0}
    \intop_{\mu_N^{\textrm{min}}}^{\mu_R-m_m}d\mu_N\acal_N(\mu_N,p_N,\rho)
       {p_{Nm}\over\mu_R}
    \end{equation}
with
\begin{equation*}
  \rho_{Nm}^0={p_{Nm}(M_R)\over M_R}.
\end{equation*}
The spectral function $\acal_N(\mu_N,p_N,\rho)$ (see Eq. (\ref{eq:spec-fct})
contains the nucleon width $\Gamma(p_N,\rho)$ in the nucleon rest frame 
(cf. Sec. \ref{sec:nucl-spec-fct}) as a function of the lab frame momentum 
$p_N$ and the density. The dependence of Eq. (\ref{eq:inmed-width})
on $p_R$  arises from the necessity of boosting the nucleon
momentum from the c.m. frame into the lab frame. The quantity 
$\mu_N^{\textrm{min}}$ is a numerical parameter for the minimal nucleon
mass considered in out calculations (see Sec. \ref{sec:buu-model}). 
It is chosen in a way that the dependence of the widths on this 
parameter is negligible.

For lower resonance momenta, Pauli blocking of outgoing nucleons is important. 
This can be accounted for by averaging over the decay angle in the 
c.m. frame \cite{effe_abs}:
\begin{equation*}
  \tilde\Gamma_{S_{11}\to Nm}^*(\mu_R,p_R,\rho)=
  {1\over 2}\int_{-1}^1 d\cos\theta\, \Theta(E-E_F(\rho))\,
   \Gamma_{Nm}^*(\mu_R,p_R,\rho).
\end{equation*}
Here, $E$ is the energy of the decaying nucleon and $E_F$ is the local
Fermi energy; in the case of offshell nucleons at $T=0$ the relevant 
information about the occupation of states in the (local) nuclear matter is 
given by the Fermi energy and not by the Fermi momentum.

\begin{figure}
\begin{center}
\includegraphics[width=15cm]{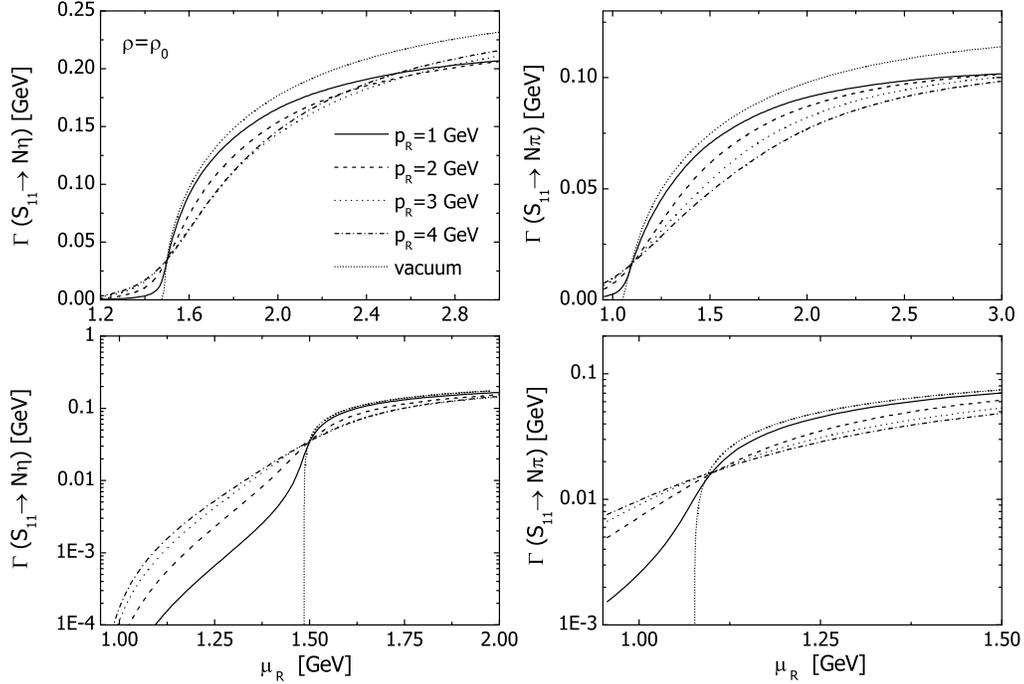}
\end{center}
\vspace{-0.5cm}
\caption{In-medium decay width of the $S_{11}(1535)$ for the channels
$N\eta$ and $N\pi$ as a function of the resonance mass. The different 
curves correspond to different resonance momenta as indicated in the
legend. The short-dotted curves show the vacuum case. The calculations were
performed at nuclear matter density $\rho_0$. Pauli blocking is not included.}
 \label{fig:s11width}
\end{figure}

In Fig. \ref{fig:s11width} we show the results of the decay widths 
$S_{11}\to N\eta$ (left panel) and $S_{11}\to N\pi$ (right panel) for density 
$\rho_0$ and different resonance momenta as a function of $\mu_R$. Pauli 
blocking is not included. The short-dotted curves show the vacuum widths. 
For the sake of a better resolution in the threshold region, the results are 
also shown in log plots in the lower part of Fig. \ref{fig:s11width}. The 
influence of the nucleon spectral function leads to a decrease of the partial 
widths for larger resonance masses and to finite, non-zero values for masses 
below the vacuum threshold. 
This behavior is the more pronounced the larger the resonance momentum is 
chosen and is strongly connected to the increase of the nucleon width with 
nucleon momentum observed in Sec. \ref{sec:nucl-spec-fct}. However, we have 
checked that the sensitivity to the components of the nucleon spectral
function with very large widths is rather small. Actually, for resonance
momenta below 4 GeV we would obtain almost the same results, when we cut the 
nucleon width at $p_N=2$ GeV in Fig. \ref{fig:nwidth-dens}, leaving the 
nucleon width constant for momenta above. This is due to the fact that
the average nucleon momenta stemming from the $S_{11}$ decay for such 
resonance momenta does not exceed $p_N\sim2$ GeV significantly in the
given resonance mass range.

\begin{figure}
\begin{center}
\includegraphics[width=15cm]{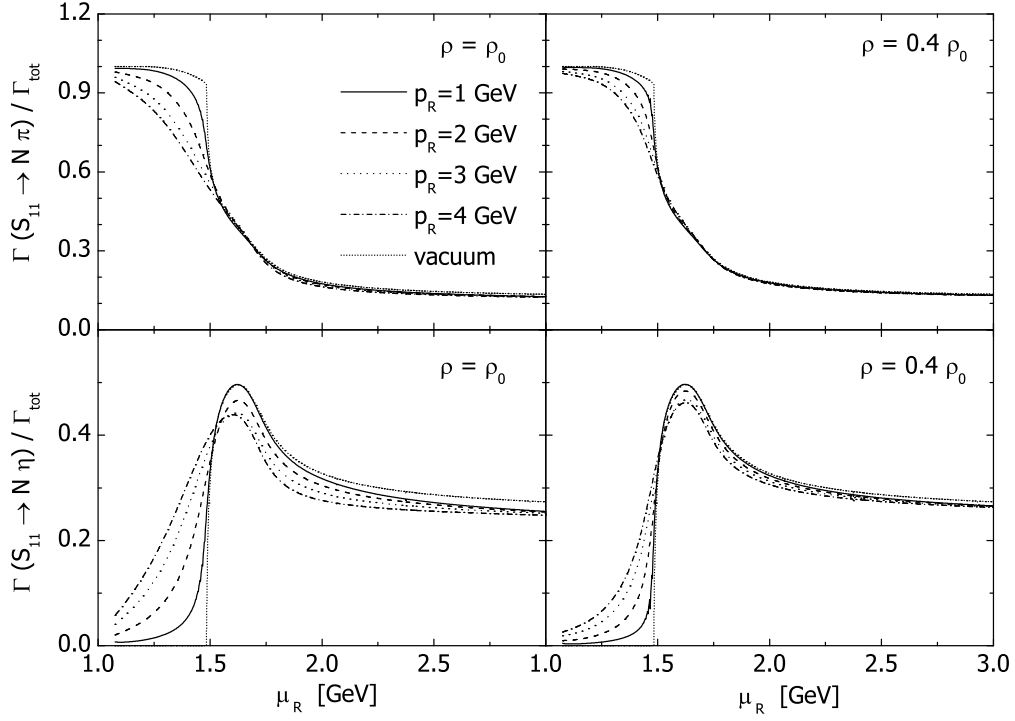}
\end{center}
\vspace{-0.5cm}
\caption{$S_{11}$ branching ratios for $N\eta$ and $N\pi$, modified with
the nucleon spectral function, as a function of the resonance mass.
The left panels show the results for density $\rho_0$, the right
panels for density $0.4\rho_0$. The different curves correspond to 
different resonance momenta as indicated in the legend, the short-dotted 
curves show the vacuum case. Pauli blocking is not taken into account.}
 \label{fig:bratio}
\end{figure}

Since we want to calculate production reactions of the type 
$\gamma N\to S_{11}\to N\eta$, it is worthwhile to see how the branching
ratios $\Gamma_{S_{11}\to N\eta,\pi}/\Gamma_{\textrm{tot}}$ are influenced.
In Fig. \ref{fig:bratio} we show the branching ratios for $N\pi$ and $N\eta$
for different momenta
as a function of the resonance mass. Since in photon-nucleus reactions also
densities smaller than $\rho_0$ are encountered, we also 
show the situation for density $\rho=0.4\rho_0$ on the right-hand side.
The short-dotted curves again show the vacuum case.
In the $N\pi$ channel we observe a step-like function that comes about
because for $m_N+m_\pi<\mu_R<\mu_N+m_\pi$ pion decay is the only relevant
decay channel. The partial pion width exhibits a strong smoothening
of the step-like structure with $p_R$ which is less pronounced
at lower density and/or lower
momentum values. The branching ratio $N\eta$ which has a clear cut in the 
vacuum at the onshell $N\eta$ threshold $m_N+m_\eta$. This step becomes 
washed out when the density and the resonance momentum increase, whereas the 
peak structure close to the $S_{11}$ resonance maximum is slightly
suppressed.

%%%%%%%%%%%%%%%%%%%%%%%%%%%%%%%%%%%%

\section{The elementary reaction}
 \label{sec:el-react}

\subsection{Kinematical situation}
 \label{sec:kin-sit}

We now discuss the sensitivity of photon- and electron-induced reactions to 
the correlated
part of the nucleon spectral function in the groundstate of the nucleus. 
In \cite{photoeta,electroeta} we have modeled
photon-nucleus and electron-nucleus reactions by the absorption of the
real or virtual photon on a single nucleon. Therefore, the quantity of 
interest is the invariant mass of the photon-nucleon pairs inside the nucleus. 

The invariant mass spectra are generated by distributing the nucleons inside
the nucleus according to a Woods-Saxon density in coordinate space. Using
local density approximation, the momenta and masses are determined according
to the spectral function discussed in Sec. \ref{sec:nucl-spec-fct}. This 
is the same procedure also used for the initialization of the nuclei for
our transport calculations, for details we therefore refer to
\cite{effe_dilep,electro}. For simplicity, the mean-field potential is 
neglected. The nucleons satisfy the 
condition $E_N\le E_F(\rho)$. Masses and momenta are independent;
the maximal possible momentum of a nucleon with mass $\mu$ is given by
$p_{\textrm{max}}^2=E_F^2-\mu^2$. For $\mu<m_N$ this value is 
larger than the local Fermi momentum $p_F=\sqrt{E_F^2-m_N^2}$, and 
according to Sec. \ref{sec:nucl-spec-fct} this is the case for most of the
nucleons off the mass shell.

\begin{figure}
\begin{center}
\includegraphics[width=15cm]{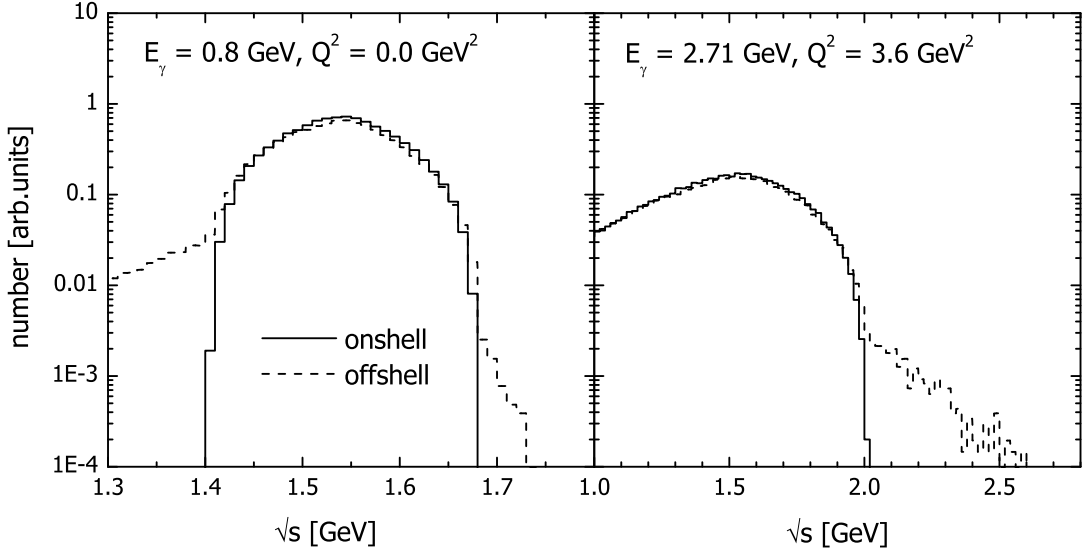}
\end{center}
\vspace{-0.5cm}
\caption{Invariant mass spectra of photon-nucleon pairs in calcium for
real and virtual photons. For both kinematics, the absorption of the
photons on an onshell nucleon at rest corresponds to an excitations of
a resonance with mass $\mu_R\sim 1.54$ GeV. The solid 
and dashed curves show results without and with nucleon spectral function
due to groundstate correlations.}
 \label{fig:srts-spec}
\end{figure}

In Fig. \ref{fig:srts-spec} we show the invariant mass spectra for a 
real photon of energy $E_\gamma=0.8$ GeV and a virtual photon with
$Q^2=3.6$ GeV$^2$ and $E_\gamma=2.71$ GeV. Both kinematics correspond to
the second resonance region; the maxima of both distributions (generated e.g.
by onshell nucleons at rest) are located at the same value of 
$\sqrt s\sim 1.54$ GeV. The solid spectra show the result using onshell 
nucleons and are limited in both cases at some maximal and minimal value of 
$\sqrt s$. It is also seen that the spectrum for virtual photons extends over 
a much larger c.m. energy range than in the real photon case. This effect is 
due to Fermi motion \cite{electro,electroeta}. Using the nucleon spectral 
function, some strength is moved to smaller and also to larger invariant 
masses, which is accompanied by a depletion in the region of the distribution 
maximum due to the depletion of the onshell peak. 

For the virtual photon case the contribution at low invariant masses extends
to $\sqrt s$ values smaller 
than 1 GeV and is not shown here. It is obvious that the contribution for
$\sqrt s$ values larger than the maximal onshell c.m. energy is more 
pronounced at finite $Q^2$. Indeed, this behavior is the stronger the
larger $Q^2$ becomes. This can be seen as follows. A measure of the
strength of this effect is the difference between the maximal 
c.m. energy values in the onshell and the offshell calculation
$\Delta s:=s_{\textrm{max}}^{\textrm{off}}-s_{\textrm{max}}^{\textrm{on}}$.
As described above, real and virtual photons probe the same kinematical
region if the c.m. energies corresponding to the maximum of the distributions,
e.g. for a  onshell nucleon at rest, are identical:
\begin{equation*}
  m_N^2+2E_\gamma m_N\overset{!}{=}m_N^2-Q^2+2E_\gamma^\prime m_N
\end{equation*}
with momenta pointing into a direction opposite to that of the photon 
momentum. Therefore, $E_\gamma<E_\gamma^\prime$ and hence 
$p_\gamma=E_\gamma<p_\gamma^\prime=\sqrt{Q^2+E_\gamma^{\prime\, 2}}$.
The maximal c.m. energies in the onshell spectra are due to
nucleons of maximal momentum (i.e. Fermi momentum $p_F$) with 
momentum vectors pointing in the opposite direction as the photon momentum
vector. Therefore,
\begin{align*}
  s_{\textrm{max}}^{\textrm{on}}(Q^2=0)&=m_N^2+2(E_\gamma E_F+
   E_\gamma p_F)\\
  s_{\textrm{max}}^{\textrm{on}}(Q^2)&=m_N^2-Q^2++2(E_\gamma^\prime E_F+
   p_\gamma^\prime p_F).
\end{align*}
In the offshell spectra, maximal c.m. energies originate from nucleons of some
mass $\mu<m_N$ and maximal possible momentum $p_{\textrm{max}}=
\sqrt{E_F^2-\mu^2}>p_F$. Here we have
\begin{align*}
  s_{\textrm{max}}^{\textrm{off}}(Q^2=0)&=\mu^2+2(E_\gamma E_F+
   E_\gamma p_{\textrm{max}})\\
  s_{\textrm{max}}^{\textrm{off}}(Q^2)&=\mu^2-Q^2+2(E_\gamma^\prime E_F+
   p_\gamma^\prime p_{\textrm{max}}).
\end{align*}
We now construct the differences
\begin{align*}
  \Delta s(Q^2=0) &= \mu^2-m_N^2+2E_\gamma(p_{\textrm{max}}-p_F)\\
  \Delta s(Q^2)   &= \mu^2-m_N^2+2\sqrt{Q^2+E_\gamma^{\prime\, 2}}
    (p_{\textrm{max}}-p_F).
\end{align*}
Since $E^\prime_\gamma
 >E_\gamma$, the effect is large when $Q^2$ becomes
large. It should be mentioned, that this also happens for real
photons with larger energy.

Due to the behavior of the c.m. spectrum, we can expect an influence of
the groundstate correlations on the threshold region of cross sections for 
particle production. Moreover, close to the maximum of the cross section
(i.e. in resonant production processes), a decrease will be observed due
to the depletion of strength close to the maximum of the $\sqrt s$ spectrum. 
Eta photo- and electroproduction therefore might be a candidate
to observe such effects, because this reaction is dominated by the
resonance $S_{11}(1535)$.

\subsection{Cross sections}
 \label{sec:xsection}

We now describe how the cross section for the reactions $\gamma A\to \eta X$ 
and $eA\to e^\prime\eta X$ is calculated. For details we refer again to
\cite{photoeta,electroeta}, where these reactions were discussed in detail.
For the elementary reactions $\gamma N$ and $\gamma^* N$ the outgoing states 
$P_{33}(1232)$, $D_{13}(1520)$, $S_{11}(1535)$, $F_{15}(1680)$, $N\pi$ and 
$N\pi\pi$ are important; the cross sections for these processes were 
discussed there both for $Q^2=0$ and 3.6 GeV$^2$. It was also mentioned that 
for $\eta$ production on nuclei the elementary reaction 
$\gamma N\to S_{11}(1535)$ with a subsequent decay into $N\eta$ is the most 
important source in the second resonance region. At finite $Q^2$, this 
dominance is somewhat reduced due to contributions from secondary processes 
in the FSI, e.g. $\gamma N\to N\pi,\pi N\to S_{11}\to N\eta$, which in 
photoproduction are kinematically suppressed. In vacuum, $\eta$ photoproduction 
data on the proton are described by the Breit-Wigner ansatz:
\begin{align}
  \label{eq:xsection-eta}
  \sigma_{\gamma N\to S_{11}\to \eta N}(\sqrt s)&=\left({k_0\over k}\right)^2
   {s\Gamma_\gamma(\sqrt s)\,\Gamma_{S_{11}\to X}(\sqrt s)\over
   (s-M_{S_{11}}^2)^2+s\Gamma_{S_{11}\to X}^2(\sqrt s)} 
   {2m_N\over M_{S_{11}}\Gamma_0}\vert A_{1/2}^p\vert^2,\nonumber\\
   &=\sigma_{\gamma N\to S_{11}}\cdot{\Gamma_{S_{11}\to \eta N}(\sqrt s)\over 
         \Gamma_{S_{11}\to X}(\sqrt s)}.
\end{align}
with $\Gamma_\gamma=\Gamma_0\cdot k/k_0$ and the
pole-mass decay width $\Gamma_0=0.151$ GeV. The c.m. momentum $k$ of the 
$\gamma p$ pair depends on the c.m. energy $\sqrt s$ (i.e. mass of the 
resonance), $k_0=k(M_{S_{11}})$ is the c.m. momentum taken at the pole mass of 
the $S_{11}$. This expression can also be used for electroproduction; in the
one-photon exchange approximation this is then the cross section for the
reaction of the exchanged virtual photon with the nucleon. The cross section
for the whole electron nucleon reaction is obtained by multiplication with
the virtual photon flux factor $\Gamma_v$ (Hand convention \cite{hand}):
\begin{equation}
 \label{eq:el-xsection-eta}
 {d\sigma_{eN\to e^\prime N\eta}\over dE^\prime d\Omega}=
  \Gamma_v\cdot\sigma_{\gamma^* N\to N\eta}\qquad
 \Gamma_v={\alpha\over 2\pi^2}{s-m_N^2\over 2m_N}{E^\prime\over Q^2 E}
   {1\over 1-\varepsilon}.
\end{equation}
In Eq. (\ref{eq:xsection-eta}), $A_{1/2}$ denotes
the transverse helicity amplitude of the $S_{11}$, which depends on $Q^2$.
In electroproduction, there is also a longitudinal contribution, which is
negligible as well as the dependence of the cross section 
(\ref{eq:el-xsection-eta}) on the polarization parameter $\varepsilon$ 
\cite{electroeta}. Background processes for both reactions are negligible
\cite{penner}.

The cross section in Eq. (\ref{eq:xsection-eta}) is a function of the 
invariant mass of the photon-nucleon pair (i.e. the resonance mass). 
Therefore, there will be an influence of the groundstate correlations 
as described in the last section. Moreover, the decay widths 
of the $S_{11}(1535)$ enter the expression as well as the branching ratio for
$N\eta$. In the onshell case, the cross section (\ref{eq:xsection-eta})
has a cut at the $N\eta$ vacuum threshold $\sqrt s=m_N+m_\eta$ due to
the branching ratio (cf. Fig. \ref{fig:bratio}). This is different when the
modified branching ratio is used: The mass spectra of the $S_{11}(1535)$
resulting from the elementary reaction, are essentially given by the
c.m. energy spectra in Fig. \ref{fig:srts-spec}. Since the modified
branching ratio assumes finite values at invariant masses below
$m_N+m_\eta$, all resonances with such small masses will yield small
but finite contributions to $\eta$ production in nuclei. This will clearly
have an impact on the final result.
This effect will be more pronounced at finite $Q^2$, because
here the sensitivity to the correlated part of the nucleon spectral
function is larger and due to the large resonance momenta also the 
modification of the branching ratio will be stronger.

We now show how the cross section for the $\gamma A\to \eta X$
reaction is calculated within our model \cite{effe_abs,effe_pion}:
\begin{equation}
  \label{eq:xsection-nucl}
  \sigma_{\gamma A\to \eta X}(E_\gamma)=g\intop_Vd^3r 
  \int{d^3p_N\over (2\pi)^3}d\mu_N\acal(\mu_N,p_N,\rho)\,
  \Theta(E_F(\rho(\vec r))-E_N)
 \sum_f\sigma_{\gamma_N\to f} M_f^\eta,
\end{equation}
where the coordinate space integral extends over the nuclear volume and
the sum runs over all possible final states of the elementary reaction
listed above. $M_f^\eta$ are multiplicity factors taking into account how 
many etas were produced by the particles contained in the final state $f$. 
For the energies under consideration, $M_f^\eta$ usually equals one or zero. 
The multiplicity factor includes the effects from the FSI and will
be determined by the BUU model, described in Sec. \ref{sec:buu-model}.
For $\eta$ electroproduction, a formula analogous to Eq. (\ref{eq:xsection-nucl})
exists for the calculation of the cross section
$d\sigma(eA\to e^\prime\eta X)/(d\Omega dE^\prime)$. The main difference is
that the elementary cross sections $\sigma_{\gamma N}$ have to be
substituted by their counterparts for the electron-nucleon reaction,
$d\sigma_{eN}/(dE^\prime d\Omega)$. More details can be found in 
\cite{electro,electroeta}.

%%%%%%%%%%%%%%%%%%%%%%

\section{The BUU model}
 \label{sec:buu-model}

For the description of the FSI we make use of the BUU transport model 
developed in \cite{teis,effe_dilep,electro}, which yields a realistic 
description of different reaction types included in the FSI as elastic and 
inelastic collisions, resonance formation and decay and particle production by 
string fragmentation in high energy events. 
The model is based upon the BUU equation, which describes the offshell
space-time evolution of the spectral phase space density  $F_i=\acal_i f_i$
of all particle species $i=N,\pi,\eta,P_{33},S_{11},...$:
\begin{equation}
  \label{eq:buu-eq}
    \left({\partial\over\partial t}+\vec\nabla_p H_i\cdot\vec\nabla_r
  -\vec\nabla_r H_i\cdot\vec\nabla_p\right)F_i(\vec r,\vec p,\mu;t)
=I_{\textrm{coll}}[F_N,F_\eta,F_{S_{11}},...].
\end{equation}
Here  $\acal_i$ is the spectral function and $H$ is a relativistic Hamilton 
function 
\begin{equation}
  \label{eq:hamilton-fkt}
  H_i=\sqrt{(m_i+S_i)^2+p^2}.
\end{equation}
In the case of baryons, $H$ includes an effective scalar mean-field potential 
$S_i$. For the present calculations, we use the same momentum dependent 
parametrization as in \cite{photoeta}.
The reaction processes in the FSI are described by a set of BUU equations in 
the following way: Whereas the propagation according to the Hamilton function 
is described by the left-hand side, the right-hand side (collision integral) 
includes a loss and a gain term, consisting of collision rates for the 
different collision processes and accounting for the gain and the loss of the
spectral phase space density $F_i$ at each point caused by the collisions.
The BUU equations are therefore coupled via the collision terms.
Pauli blocking is explicitly taken into account in each collision
reaction with outgoing nucleons. Therefore, the $S_{11}$ widths without
Pauli blocking have to be used when determining whether a resonance
decays or not in order to avoid double counting.

The common method to solve this system of equations is the so-called
test particle ansatz. The spectral phase space density $F_i$ is substituted by
\begin{equation}
  \label{eq:test-part}
  F_i(\vec r,\vec p,\mu;t)\propto
\sum_i\delta(\vec r-\vec r_i(t))\delta(\vec p-\vec p_i(t))
   \delta(\mu-\mu_i(t)).
\end{equation}
In the absence of collisions (i.e. vanishing collision integral), the BUU 
equation is fulfilled by this ansatz if the test particles obey certain 
equations of motion. In the onshell approximation, these are the Hamiltonian 
equations of motion:
\begin{equation*}
  \dot{\vec r}={\partial H\over \partial\vec p},\qquad
  \dot{\vec p}=-{\partial H\over \partial\vec r}.
\end{equation*}
For the offshell propagation of the nucleons, we use the method
developed in \cite{effe_offshell}. There a scalar 'offshell potential'
was introduced by
\begin{equation*}
  \Delta\mu(t)=(\mu_N(t_{\textrm{cr}})-m_N){\rho(t)\over \rho(t_{\textrm{cr}})}
\end{equation*}
with the baryon density $\rho$ and the creation time $t_{\textrm{cr}}$. The 
full Hamilton function for the nucleons then reads
\begin{equation*}
  H=\sqrt{(m_N+U_S+\Delta\mu)^2+p^2}.
\end{equation*}
This ansatz ensures that the offshellness vanishes when the nucleon test
particles travel into the vacuum. For the propagation, the usual
equations of motion are used with the Hamilton function including the
offshell potential. The time evolution of the offshell mass is directly given
by 
\begin{equation*}
  \mu_N(t)=m_N+\Delta\mu(t)=m_N+
  (\mu_N(t_{\textrm{cr}})-m_N){\rho(t)\over \rho(t_{\textrm{cr}})},
\end{equation*}
which means that the nucleon mass is shifted towards the pole mass $m_N$
when travelling to regions of small density.
This equation of motion does not prevent the mass from becoming negative
during the propagation. In order to avoid such effects, we choose a
minimal nucleon mass $\mu_N^{\textrm{min}}=0.4$ GeV in the initialization
and collisions. We have checked that this value has no influence on the
final results. For consistency, we use this value also in the calculation
of the modified $S_{11}$ decay width in Eq. (\ref{eq:inmed-width}).

This set of equations has found its a posteriori justification in recent
work in which test particle equations of motion were derived from the
Kadanoff-Baym equation using a gradient expansion \cite{leupold,cassjuch}.
The equations derived there become equal to the ones used here in the limit
of not too large offshellness and $\Gamma\sim\rho$.

%%%%%%%%%%%%%%%%

\section{Results}
 \label{sec:results}

\begin{figure}
\begin{center}
\includegraphics[width=12cm]{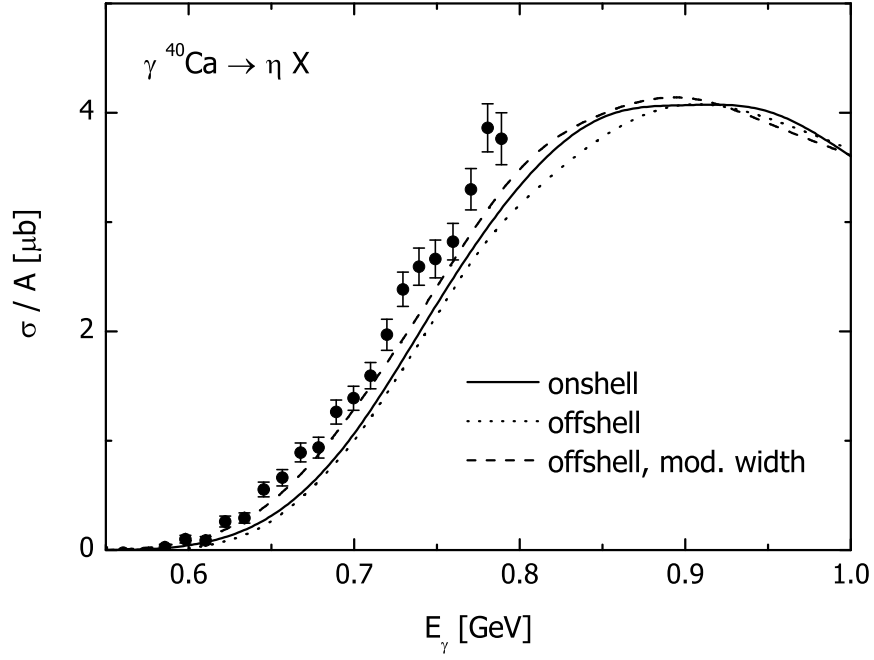}
\end{center}
\vspace{-0.5cm}
\caption{Results for the reaction $\gamma \textrm{Ca}\to \eta X$. The curves
correspond to calculations with onshell nucleons (solid), offshell
nucleons (dotted) and offshell nucleons with modified $S_{11}$ widths
(dashed). The data are from \cite{eta_roebig}.}
 \label{fig:gamca-eta-mod}
\end{figure}

In Fig. \ref{fig:gamca-eta-mod} we show the cross section for the reaction 
$\gamma \textrm{Ca}\to \eta X$, determined according to
Eq. (\ref{eq:xsection-nucl}). The solid curve shows the result with
onshell nucleons. The dotted line is the offshell result but using vacuum
widths for the elementary cross sections involving the $S_{11}$ (Eq. 
(\ref{eq:xsection-eta})). 
We see that the curve is lowered compared to the onshell calculation,
especially in the region close to the maximum. This is essentially the
effect of the groundstate correlations and originates in the depletion in the 
c.m. energy spectra close to the maxima, as discussed in Sec. 
\ref{sec:kin-sit}. 
Moreover, no 
effects in the region close to threshold are observed, as anticipated before.
Since the the absolute threshold for (coherent) $\eta$ production on nuclei is 
located at $E_\gamma\sim0.55$ GeV, there are no contributions below the
onshell threshold. The dashed curve finally shows the result including both the
nucleon spectral function and the modified decay widths for the $S_{11}$.
Compared to the dotted offshell curve, an increase in the energy region
from threshold to the maximum can be seen. This is the effect of the 
width modification. However, the net increase (i.e. compared to the onshell 
result) for energies larger than 0.8 GeV is essentially zero, because the
increase caused by the modified widths is canceled by the 
decrease due to the groundstate correlations. Therefore, we are left only 
with a slight increase in the threshold region with respect to the onshell
calculation, yielding a closer agreement with the experimental data from 
\cite{eta_roebig}.

\begin{figure}
\begin{center}
\includegraphics[width=10cm]{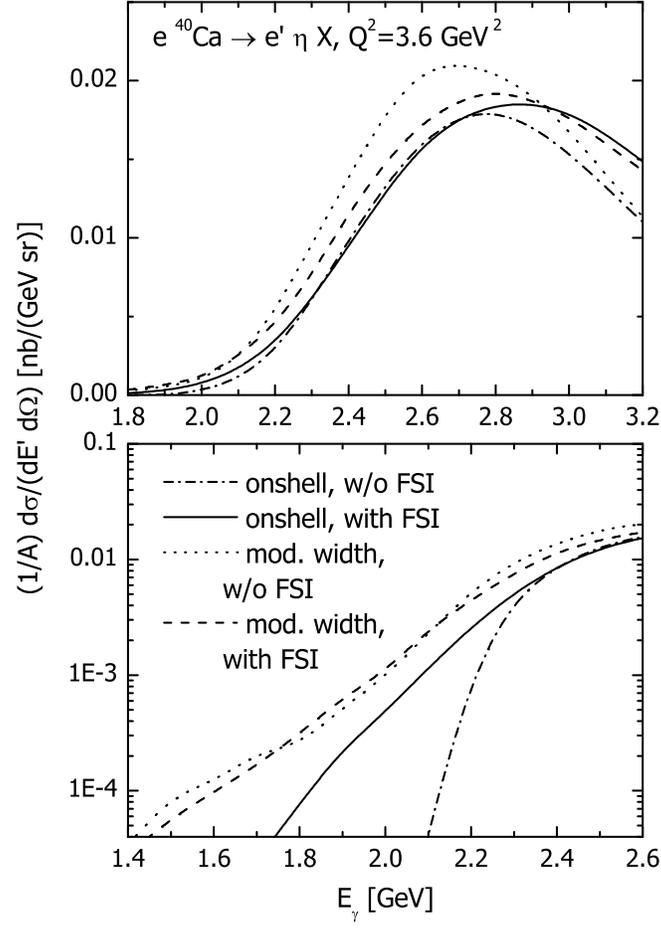}
\end{center}
\vspace{-0.5cm}
\caption{Cross section for the reaction $e\textrm{Ca}\to e^\prime\eta X$.
The solid and dash-dotted curves show the onshell calculation with and
without FSI, the dashed and dotted lines show the results with offshell
nucleons and modified $S_{11}$ width, with and without FSI.}
 \label{fig:elca-eta-mod}
\end{figure}

We now turn to $\eta$ electroproduction. Here we expect a somewhat larger
effect: On the one hand, the virtual photon is more sensitive to the
large momentum component of the spectral function of correlated nucleons
in the elementary reaction. On the other hand, the $S_{11}$ are produced
with larger momenta, which increases the medium effect of the decay
widths. In Fig. \ref{fig:elca-eta-mod} we show several scenarios for 
the reaction $e\textrm{Ca}\to e^\prime\eta X$. The lower plot shows the
threshold region in more detail. The solid and dash-dotted 
curves show the onshell results with and without FSI. It is seen that
the calculation with FSI yields a somewhat larger cross section. This effect, 
discussed in \cite{electroeta}, is due to the fact that in reactions with 
large momentum transfer a large contribution of etas is produced in secondary
processes in the FSI. Therefore, the absorption of etas from the primary
production chain $\gamma N\to S_{11}\to \eta N$ is balanced. 
For small photon energies $E_\gamma<2.3$ GeV this is even more obvious.
Due to the secondary contributions, which are not present in a calculation 
without FSI, the threshold is lowered down to energies of about 1.8 GeV. It
should be mentioned, that the absolute (coherent) threshold for this reaction 
is given by $E_\gamma=m_\eta^2+(m_\eta^2+Q^2)/(2m_A)\sim 0.6$ GeV is located
at much smaller photon energies.
The dashed and dotted lines present the offshell calculation with and 
without FSI, including the modified $S_{11}$ decay widths.
The calculation without FSI is enhanced compared to the onshell result
by up to $\sim20$\% close to the maximum of the curve. Similar to
photoproduction, this increase is due to the width modification.
As mentioned in Sec. \ref{sec:xsection}, also resonances with masses
below the free $N\eta$ threshold contribute now, this effect is stronger for 
large $Q^2$.
Moreover, there is a strong subthreshold effect caused by the groundstate 
correlations. In the full calculation with FSI, the difference to the onshell
counterpart becomes somewhat smaller. This is mainly due to the contribution
of secondary processes at higher energies (yielding approximately one half
of all etas \cite{electroeta}), which are
essentially identical for the onshell and the offshell case. However,
the remaining increase is larger than in photoproduction.
In the threshold region, the curves with and without FSI are close to
each other, because there only etas from the primary reaction influenced
by the groundstate correlations contribute. The difference to the solid
onshell curve, however, is smaller than in the scenario without FSI,
because the onshell threshold is strongly lowered by the FSI.

%%%%%%%%%%%%%%%%%%%%%%%%

\section{Summary}

We have calculated $\eta$ photo- and electroproduction on nuclei within 
the BUU transport model. In contrast to \cite{photoeta,electroeta}, we
take into account that the nucleons are correlated in the nuclear
groundstate and acquire a finite width due to collision reactions during
the FSI. This has an influence on the kinematics of the elementary 
photon-nucleon reaction. Furthermore, we have discussed how the possibility
of a resonance decay into broad nucleons might modify the in-medium
decay width of the $S_{11}(1535)$. Here we found strong deviations from
the vacuum case in particular for large resonance momenta. Combining 
all these issues together with the model for the offshell propagation
developed in \cite{effe_offshell}, we showed that there is indeed an
influence on the $\eta$ production cross section on nuclei, which is 
more pronounced at large momentum transfer, i.e. in electron induced
reactions at large $Q^2$. Besides contributions below the threshold of
the onshell calculations, an increase of the cross section for energies
up to the maximum is observed.

The model presented here is particularly suitable for an investigation of
offshell effects on incoherent particle production yields. In particular,
it could also be applied to reactions with high energy photons such as 
$J/\psi$ production on nuclei, where searches for color transparency 
phenomena have to rely on an accurate description of more involved, but
nevertheless interesting effects.

\section*{ACKNOWLEDGMENTS}

We acknowledge many valuable discussions with M. Post.
This work was supported by DFG.

% ----------------------------------------------------------------

%%%%%%%%%%%%%%%%%%%%%%%%%%%%%%%%%%%%%%%%%%%%%%%%%%%%%%%%%%%%%%%%%%%%
\newpage

%%%%%%%%%FIG 0%%%%%%%%%%%%%%%%%%%%%%
%%%%%%%%%%%%%%%%%%%%%%%%%%%%%%%%%%%%

%%%%%%%%%FIG 1%%%%%%%%%%%%%%%%%%%%%%
%%%%%%%%%%%%%%%%%%%%%%%%%%%%%%%%%%%%

%%%%%%%%%FIG 2%%%%%%%%%%%%%%%%%%%%%%
%%%%%%%%%%%%%%%%%%%%%%%%%%%%%%%%%%%%

%%%%%%%%%FIG 3%%%%%%%%%%%%%%%%%%%%%%
%%%%%%%%%%%%%%%%%%%%%%%%%%%%%%%%%%%%

%%%%%%%%%FIG 4%%%%%%%%%%%%%%%%%%%%%%
%%%%%%%%%%%%%%%%%%%%%%%%%%%%%%%%%%%%

%%%%%%%%%FIG 5%%%%%%%%%%%%%%%%%%%%%%
%%%%%%%%%%%%%%%%%%%%%%%%%%%%%%%%%%%%

%%%%%%%%%FIG 6%%%%%%%%%%%%%%%%%%%%%%
%%%%%%%%%%%%%%%%%%%%%%%%%%%%%%%%%%%%

%%%%%%%%%FIG 7%%%%%%%%%%%%%%%%%%%%%%
%%%%%%%%%%%%%%%%%%%%%%%%%%%%%%%%%%%%

\end{document}